\documentclass[a4paper]{article}
\usepackage{graphicx,amssymb}
\begin{document}

\title{Life inside black holes}
\author{V. I. Dokuchaev\thanks{e-mail: dokuchaev@inr.ac.ru}
 \\ {\it \small Institute for Nuclear Research of the Russian Academy of Sciences}}

\date{}
\maketitle

\begin{abstract}
We consider test planet and photon orbits of the third kind inside a black hole, which are
stable, periodic and neither come out of the black hole nor terminate at the singularity. Interiors of supermassive black holes may be inhabited by advanced civilizations living on planets with the third-kind orbits. In principle, one can get information from the interiors of black holes by observing their white hole counterparts.
\end{abstract}

\begin{figure}[t]
\begin{center}
\includegraphics[angle=0,width=0.4\textwidth]{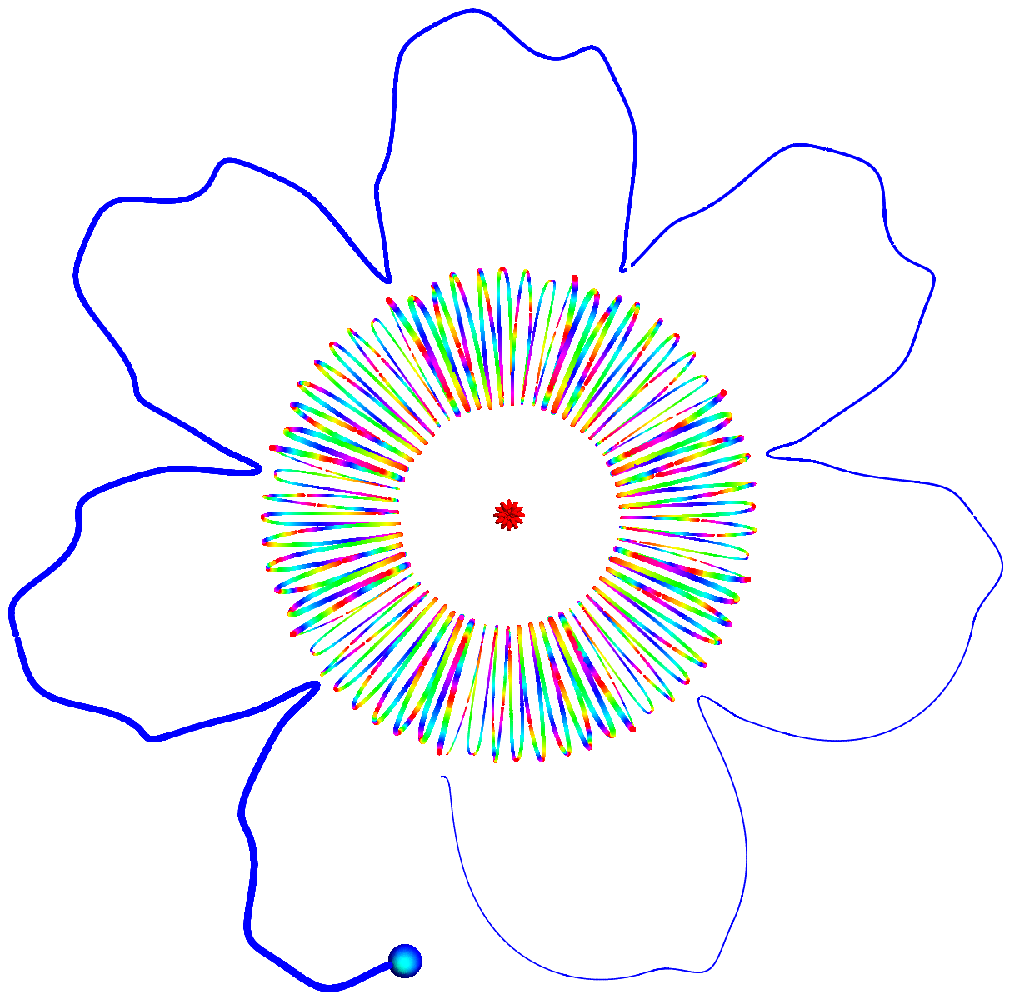}
\includegraphics[angle=0,width=0.58\textwidth]{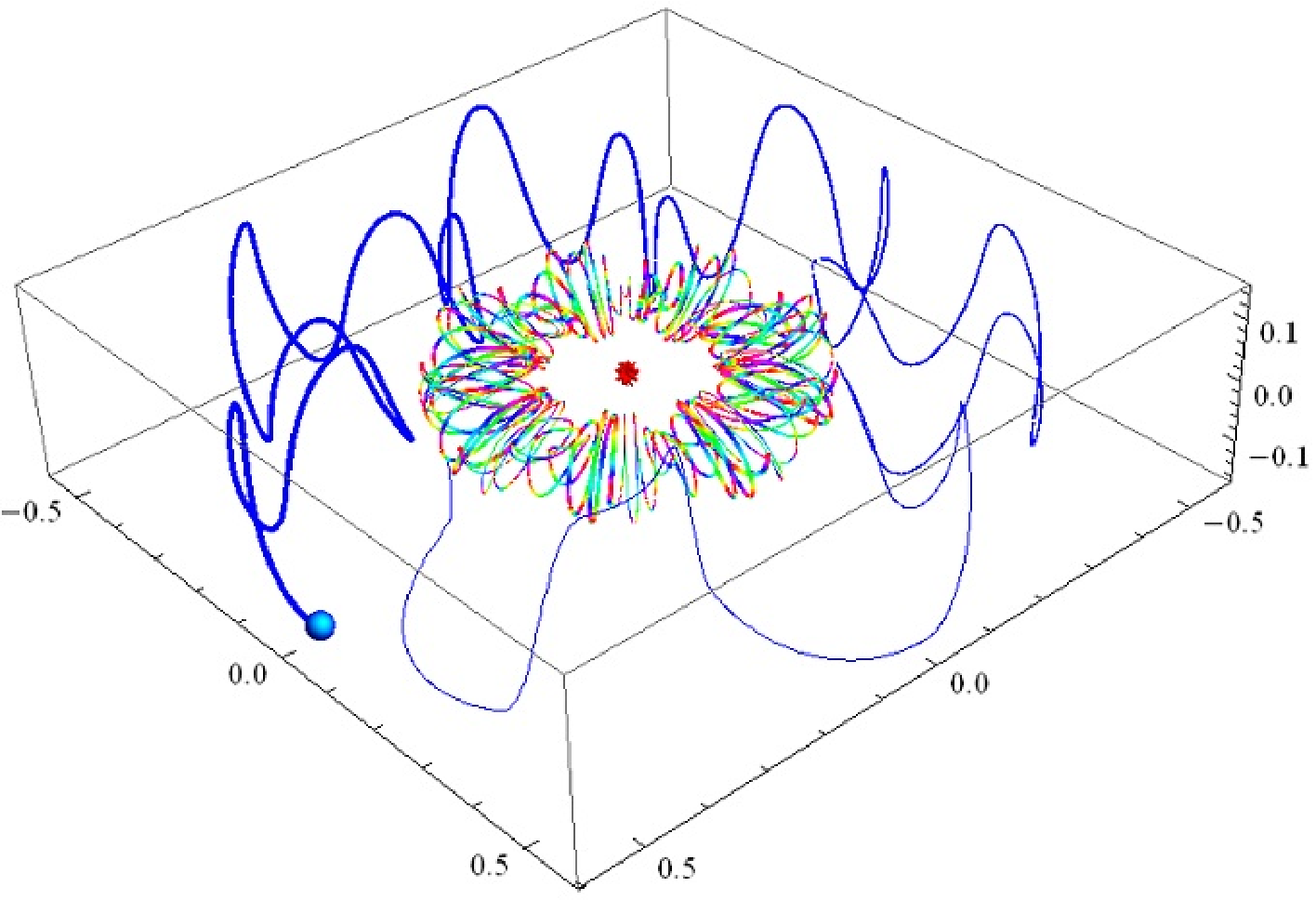}
\end{center}
\caption{A nonequatorial stable periodic orbit of a planet
(the external curve, with $E=0.568$, $L=1.13$, $Q=0.13$) and a photon orbit (the internal curve, with $b=L/E=1.38$,
$q=Q/E^2=0.03$) inside a black hole ($a=0.9982$, $e=0.05$) in the
locally nonrotating frame \cite{bpt72}, viewed from the north pole and from the
outside.}
 \label{gr1}
\end{figure}

Orbits of the third kind were described in 
\cite{Bicak89a,Bicak89b,Kagramanova09,Olevares11,Kagramanova10}
under the assumption of the Kerr-Newman metric validity inside a
black hole event horizon. The motion of a test particle (e.\,g., a
planet) with mass $\mu$ and electric charge $\epsilon$ in the
background gravitational field of a Kerr-Newman black hole (BH) with mass
$M$, angular momentum $J=Ma$ and electric charge $e$ is completely
defined by three integrals of motion: the total particle energy $E$,
the azimuthal component of the angular momentum $L$ and the Carter
constant $Q$, related to the total angular momentum of the particle.
An orbital trajectory of a test planet is governed in the
Boyer-Lindquist coordinates $(t,r,\theta,\varphi)$ by the equations of
motion \cite{Carter68,bpt72}:
\begin{eqnarray}
 \rho^2\frac{dr}{d\lambda} &=& \pm \sqrt{V_r}, \quad
 \rho^2\frac{d\theta}{d\lambda} = \pm\sqrt{V_\theta}, \label{rmot} \\
 \rho^2\frac{d\varphi}{d\lambda} &=& L\sin^{-2}\theta+a(\Delta^{-1}P-E),
 \label{phimot} \\
 \rho^2\frac{dt}{d\lambda} &=& a(L-aE\sin^{2}\theta)+(r^2\!+\!a^2)\Delta^{-1}P,
  \label{tmot}
\end{eqnarray}
where $\lambda=\tau/\mu$, $\tau$ --- is the proper time of a particle
and
\begin{eqnarray}
 V_r &=& P^2-\Delta[\mu^2r^2+(L-aE)^2+Q], \label{Vr} \\
 V_\theta &=& Q-\cos^2\theta[a^2(\mu^2-E^2)+L^2\sin^{-2}\theta],
 \label{Vtheta} \\
 P &=& E(r^2+a^2)+\epsilon e r -a L, \quad
 \rho^2 = r^2+a^2\cos^2\theta, \quad
 \Delta = r^2-2r+a^2+e^2.  \label{Delta}
 \end{eqnarray}
We use the normalized dimensionless variables and parameters:
$r\Rightarrow r/M$, $a\Rightarrow a/M$, $e\Rightarrow e/M$,
$\epsilon\Rightarrow \epsilon/\mu$, $E\Rightarrow E/\mu$,
$L\Rightarrow L/(M\mu)$, $Q\Rightarrow Q/(M^2\mu^2)$. The effective
potentials $V_r$ and $V_\theta$ in (\ref{Vr}) and (\ref{Vtheta})
determine the motion of particles in $r$- and $\theta$-directions
\cite{bpt72}. The Fig.~\ref{gr1} presents examples of third kind
nonequatorial orbits of a test planet and a photon, calculated by
numerical integration of Eqs. (\ref{rmot}) -- (\ref{tmot}).

For circular orbits of test particles with $r=const$, Eqs.
(\ref{Vr}) and (\ref{Vtheta}) provide the conditions:
\begin{equation}
 \label{circularorb}
V_r(r)=0, \quad V_r'(r)\equiv\frac{dV_r}{dr}=0.
\end{equation}
The circular orbits would be stable if $V_r''<0$, i.~e. at the
maximum of the effective potential. In the case of a rotating BH
(with $a\neq0$), a particle in an orbit with $r=const$ may be
additionally moving in the latitudinal $\theta$-direction, if
$Q\neq0$. These nonequatorial orbits are called {\em spherical
orbits} \cite{Wilkins72}. Purely circular orbits correspond to
the particular case of spherical orbits with the parameter $Q=0$. These
circular orbits are completely confined in the BH equatorial plane.
In general, there are four possible solutions (some of them may be
unstable) of Eqs. (\ref{circularorb}) for the azimuthal
momentum $L_i$ and the total energy $E_i$ of test particles with a
charge $\epsilon$ on the spherical orbits with $r=const$:
\begin{equation}
 \label{Li}
L_i=-\frac{1}{2}\left(\chi_0\pm\chi_{1,2}
 +\frac{1}{2}\frac{\eta_1}{\kappa_1}\right), \quad
 E_i=\frac{\alpha_1}{\alpha_2} +\left(\alpha_3 +
\alpha_4L_i + \alpha_5L_i^2\right)\frac{L_i}{\alpha_6},
% \label{Ei}
\end{equation}
with $i=1,2,3,4$. The expressions for coefficients $\chi$, $\eta$,
$\kappa$ and $\alpha$ in (\ref{Li}) are rather cumbersome:
\begin{eqnarray}
 \label{chi}
\chi_0&=&\sqrt{\left(\frac{1}{4}\frac{\eta_1^2}{\kappa_1}
\!-\!\frac{2}{3}\xi_1 \!+\!
 \frac{2^{1/3}}{3}\frac{\xi_4}{\xi_6} \!+\! \frac{1}{3\, 2^{1/3}}\xi_6\right)
 \frac{1}{\kappa_1}}, \\
\chi_{1,2}&=& \sqrt{\frac{1}{\kappa_1}\left[
\frac{1}{2}\frac{\eta_1^2}{\kappa_1}\!-\!\frac{4}{3} \xi_1 \!-\!
\frac{2^{1/3}}{3} \frac{\xi_4}{\xi_6} \!-\! \frac{1}{3\, 2^{1/3}}
\xi_6 \!\pm\! \frac{1}{\chi_0} \left(2 \xi_2 \!+\!
\frac{1}{4}\frac{\eta_1^3}{\kappa_1^2} \!-\! \frac{\xi_1\,
\eta_1}{\kappa_1}\right)\right]}, \\
\eta_1&=& 8 a e \epsilon x^5 \{6 e^4 \!+\! e^2 x (7 x\!-\!17) \!+\!
 a^2 [6 e^2 \!+\! (x\!-\!5) x] \!+\! x^2 [12 \!+\! x (3 x\!-\!11)]\}, \\
\kappa_1&=&4 x^6 \{4 a^2 (e^2 \!-\! x) \!+\! [2 e^2 \!+\! (x\!-\!3) x]^2\}, \\
\kappa_2&=&4 \bigl\{a^4 [Q\!+\!(e^2\!+\!Q) x\!-\!x^2]\!+\!x^3 [Q(2 e^2 \!-\!3x)\!+\!(e^2\!+\!Q) x^2\!-\!x^3] \!+\! a^2 x \{e^4\!+\!2 e^2 [Q \nonumber \\
&\!+\!& (x\!-\!2) x]
 \!-\! 2 x [Q\!-\!(2\!+\!Q) x\!+\!x^2]\}\bigr\}^2 \!+\! 4 e^2 \epsilon^2 \Bigl\{a^8 (Q \!-\! x^2) (e^2 \!+\! x^2) \!-\! x^8 [e^2 \nonumber \\
 &\!+\!& (x\!-\!2) x] (Q \!+\! x^2)
 \!+\!
 a^4 x^2 \bigl\{e^6 (\epsilon^2\!-\!2)\!-\!2 x^2 \{Q [2 \!+\! (5 \!-\! 2 x) x] \!+\! (x\!-\!1) x [4 \!+\! 3 (x\!-\!2) x]\} \nonumber \\
 &\!+\!&
 e^2 x^2 \{8 Q \!-\! 2 [8 \!+\! x (5 x\!-\!12)] \!+\! [1 \!+\! (x\!-\!4) x]
 \epsilon^2\} \!+\! 2 e^4\{Q \!+\! x [5 \!-\! 3 x \!+\! (x\!-\!1) \epsilon^2]\}\bigr\} \nonumber \\
 &\!+\!&
 a^6 \bigl\{Q (x\!-\!1 \!-\! 2 x^2)\!-\!2 x^2 \{x^2 [1 \!+\! 2 (x\!-\!2) x]\} \!+\! e^4 [Q \!+\! x^2 (\epsilon^2\!-\!3)] \!+\! e^2 x \{Q (4 x\!-\!2) \nonumber \\
 &\!+\!& x^2 [8 \!+\! x (\epsilon^2\!-\!6)]\}\bigr\} \!-\!
 a^2 x^4 \bigl\{2 x^2 \{Q (3 x\!-\!5) \!+\! x^2 [5 \!+\! 2 (x\!-\!3) x]\}
 \nonumber \\
 &\!+\!& e^4 (\!-\!5 Q  \!+\! x^2 (1 \!+\! \epsilon^2))
 \!+\! e^2 x \{2Q (7 \!-\! 2 x) \!+\! x^2 [x (6 \!+\! \epsilon^2)\!-\!2 (4\!+\!\epsilon^2)]\}\bigr\}\Bigr\}, \\
 \xi_1&=&4 x^2 \Bigl\{4 a^6 Q (e^2 \!-\! x) \!+\! 2 x^4 [2 e^2 \!+\!
(x\!-\!3) x] [2 e^2 Q \!-\! 3 Q x \!+\! (e^2 \!+\! Q) x^2 \!-\! x^3]
\!-\! e^2 x^6 [e^2 \nonumber \\
 &\!+\!& (x\!-\!2) x] \epsilon^2 +  a^4 \bigr\{\!-\!2 x^2 [x^2 (5 \!+\! x)\!-\!Q (x\!-\!5) (x\!-\!1)]\!+\!  e^4 [4 Q \!+\! x^2 (13 \epsilon^2\!-\!4)] \nonumber \\
  &\!+\!&
 e^2 x \{12 Q (x\!-\!1) \!+\! x^2 [2 (7 \!+\! x)
  \!+\!  (5 x\!-\!8) \epsilon^2]\}\bigr\}
 a^2 x^2 \bigl\{4 x^2 \{Q [3 \!+\!(x\!-\!5)x]   \nonumber \\
 &\!-\!& x[\!-\!6\!+\!x(3\!+\!x)]\}\!+\! e^6 (13 \epsilon^2\!-\!4) \!+\! 2 e^2 x \{2 x (x\!-\!2) (5 \!+\! x)
 \nonumber \\
  &\!+\!&
 2 Q (4 x\!-\!5) \!+\! x [11 \!+\! 5 (x\!-\!3) x] \epsilon^2\}
 \!+\!  2 e^4 \{4 Q \!+\! x [11 \!-\! x \!+\! (9 x\!-\!17)
 \epsilon^2]\}\bigr\}\Bigr\}, \\
\xi_2&=&8 a e x \epsilon \biggl\{a^6 \{e^2 (2 Q \!-\! x^2) \!+\! x
[Q (x\!-\!1) \!+\! x^2]\} \!+\! x^4 \bigr\{x^2 \{x^2 (\!-\!4 \!+\!
3 x) \!+\! Q [12 \!+\! x (3 x\!-\!11)]\} \nonumber \\
 &+&  e^4 (6 Q \!-\! x^2 (\epsilon^2\!-1\!)) \!+\!
 e^2 x \{Q (7 x\!-\!17) \!-\! x^2 [x\!-\!1 \!+\! (x\!-\!2) \epsilon^2]\}\bigr\}
+ a^4 \bigr\{x^2 \{x^2 (\!-\!8 \!+\! 5 x) \nonumber \\
 &\!+\!& Q [4 \!+\! 5 (\!-\!1 \!+\! x) x]\} \!+\! e^4 [2 Q \!+\! x^2 (\!-\!4 \!+\! 3 \epsilon^2)]+ e^2 x \{Q (7 x\!-\!5) \!+\! x^2 [13 \!-\! 3 x \!+\! (2 x\!-\!1)
 \epsilon^2]\}\bigr\}  \nonumber \\
&\!+\!&
 a^2 x^2 \Bigl\{x^3 [16 \!+\! x (7 x\!-\!20) \!+\! Q (7 x\!-\!15)] \!+\!
 3 e^6 (\epsilon^2\!-\!1)+ e^4 \{4 Q \!+\! x [16 \!-\! 5 x \!+\! (5 x\!-\!7) \epsilon^2]\} \nonumber \\
&\!+\!& e^2 x \bigl\{6 Q (2 x\!-\!1) \!+\! x \{22 x \!-\! 3
x^2\!-\!28
\!+\! [4 \!+\! 3 (x\!-\!3) x] \epsilon^2\}\bigr\}\Bigr\}\biggr\}, \\
\xi_3&=& 2 \xi_1^3-9 \xi_2\, \xi_1\, \eta_1+27 \kappa_2\,
\eta_1^2+27\xi_2^2\, \kappa_1-72
\kappa_2\, \xi_1\, \kappa_1, \\
\xi_4 &=& \xi_1^2-3\xi_2\, \eta_1+12\kappa_2\,\kappa_1, \quad \xi_5
= \xi_3^2-4\xi_4^3, \quad
\xi_6 = (\xi_3+\sqrt{\xi_3^2-4\xi_4^3})^{1/3}, \\
\alpha_1&=& 8 e x^2 \epsilon \Biggl\{4 a^{10} (e^2 \!-\! x)^2 (Q
\!-\! x^2) \!+\! x^{10} (2 e^2 \!+\! (x\!-\!3) x) (x (3 Q \!-\! Q x
\!+\! x^2)
\!-\! e^2 (2 Q \!+\! x^2 \nonumber \\
&\!-\!& x^2 \epsilon^2)) \!+\! a^8 \Bigl\{x^3 \bigl\{x^2 [24 \!+\!
(x\!-\!27) x] \!-\!
 Q \{8 \!+\! x [x (6 \!+\! x)\!-\!19]\}\bigr\} \!+\!
 e^4 x [4 Q (5 x\!-\!3) \!+\! x^2 (39 \nonumber \\
 &\!-\!& 21 x \!-\! 8 \epsilon^2)]
\!+\! e^2 x^2 \{3 Q [5 \!+\! (x\!-\!14) x] \!-\!
 x^2 [55 \!+\! (x\!-\!48) x \!-\! 4 \epsilon^2]\} \!+\!
 4 e^6 [Q \!+\! x^2 (\epsilon^2\!-\!2)]\Bigr\} \nonumber \\
 &+& a^6 x^2\Bigl\{4 x^3 \bigl\{(x\!-\!1) x [8 \!+\! (x\!-\!18) x] - Q \{3 \!+\! x [x (4 \!+\! x)\!-\!16]\}\bigr\} \!+\! 4 e^8
(\epsilon^2\!-\!1) \nonumber \\
&\!+\!&  e^4 x \{Q (44 x\!-\!56)
 \!-\! 4 x [17 \!+\! 11 (x\!-\!3) x] \!+\!
 x [19 \!+\! (x\!-\!24) x] \epsilon^2\}
 \!+\! e^6 \{20 Q\! \!+\!\! x [27\! -\! 23 x \nonumber \\
 &\! \!+\!\!& 3 (3 x\!-\!5) \epsilon^2]\}\!
 + e^2 x^2 \bigl\{Q [47\! \!+\!\! x (7 x\!-\!110)]\! \!+\!\!
 x \{76 \!-\! 215 x \!+\! 115 x^2 \!-\! 4 x^3 \!-\! [8 \!+\! (x \nonumber \\
 &\!-\!& 15) x]
 \epsilon^2\}\bigr\}\Bigr\} \!+\! a^2 x^6
 \bigl\{4 x^3 [ Q(9 x\!-\!27) \!+\! 2 (3 \!+\! Q) x^2 \!-\! (10 \!+\! Q) x^3 \!+\!
 x^4] \!+\! 4 e^6 [9 Q \!-\! 2 x^2 (\epsilon^2 \nonumber \\
 &\!-\!& 1)]   \!-\!  2 e^4 x \{\!-\!6 Q (x\!-\!13)
 \!+\! x^2 [7 \!+\! 9 x \!+\! (x\!-\!13) \epsilon^2]\} \!+\!
 e^2 x^2 Q [225 \!-\! 7 x (6 \!+\! x)] \!+\!
 x^2 \{\!-\!13  \nonumber \\
 &\!+\!& 53 x \!-\! 4 x^2 \!+\! [x (5 \!+\! 2 x)\!-\!21] \epsilon^2\}\bigr\}
 \!+\! a^4 x^4 \biggl\{2 x^3 \bigl\{x \{x [76\!+\!3 (x-15) x]-24\}
 \nonumber \\
 &-& Q (x-3) [x (13\!+\!3 x)-8]\bigr\} \!+\! 6 q^8 (\epsilon^2-1)
 \!+\! q^6 \{44 Q\!+\!x [41\!-\!13 x\!+\!(x\!-\!23) \epsilon^2]\} \nonumber \\
 &\!-\!& q^4 x \bigl\{Q (152\!-\!44 x) \!+\! x \{104\!-\!115 x\!+\!43 x^2
 \!+\![x (10\!+\!x)\!-\!29] \epsilon^2\}\bigr\}
 \!+\! q^2 x^2 \Bigl\{Q [161 \nonumber \\
 &\!+\!& (x\!-\!130) x]
 \!+\! x \bigl\{116\!-\!245 x\!+\!123 x^2\!-\!6 x^3\!+\!\{x [14\!+\!x (3\!+\!x)]\!-\!12\} \epsilon^2\bigr\}\Bigr\}\biggr\}\Biggr\}, \\
\alpha_2&=& 32 \biggl\{\{x^4 \!+\! a^2 [x (2 \!+\! x)\!-\!e^2]\}
\Bigl\{a^2 x^2 [a^2 (x\!-\!e^2) \!+\! x^2 (3 x\!-\!2 e^2)]^2
 \{Q (x\!-\!1)  \nonumber \\
 &\!+\!&  x [a^2 \!+\! e^2 \!+\! x (2 x\!-\!3)]\}
 \!+\! a^2 [2 x^3 \!+\! a^2 (1 \!+\! x)]
 \{x^4 \!+\! a^2 [x (2 \!+\! x)\!-\!e^2]\}
 \{e^2 x^2 \epsilon^2\!-\![a^2 \!+\! e^2  \nonumber \\
 &\!+\!& (x \!-\! 2) x] (Q \!+\! x^2)\}
 \!-\! [2 x^3 \!+\! a^2 (1 \!+\! x)]^2
 \{2 a^2 (e^2 \!-\! 2 x) [a^2 \!+\! e^2
 \!+\! (x\!-\!2) x] (Q \!+\! x^2)  \nonumber \\
 &\!-\!& e^2 x^2 [(x\!-\!1 \!) x^4 \!+\! a^4 (1 \!+\! x) \!+\!
 2 a^2 (e^2 \!-\! 2 x \!+\! x^3)] \epsilon^2\}\Bigr\} \!+\! [a^2 \!+\! e^2
 \!+\! (x\!-\!2) x] [2 x^3   \nonumber \\
 &\!+\!& a^2 (1 \!+\! x)]^3
 \bigl\{e^2 x^2 \{a^2 [e^2 \!-\! x (2 \!+\! x)]\!-\!x^4\} \epsilon^2
 \!-\![a^2 (e^2 \!-\! 2 x)^2 (Q \!+\! x^2)]\bigr\}\biggr\}, \\
\alpha_3&=& \!-\!16 a x^2 \Biggl\{2 a^8 (e^2 \!-\! x)^2 [Q(1\!-\! x)
\!+\! (e^2 \!-\! x) x] \!+\! x^7 [2 e^2 \!+\! (x\!-\!3) x]
 \{(3 x\!-\!2 e^2) [2 e^2 Q \nonumber \\
 &\!-\!& 3 Q x  \!+\! (e^2 \!+\! Q) x^2 \!-\!
 x^3] \!+\! 2 e^2 x^2 [2 e^2 \!+\! x (4 x\!-\!3)] \epsilon^2\} \!+\!
 a^6 x %\bigl\{
 (e^2 \!-\! x) \biggl\{2 e^6 \!+\! 2 e^4 x (7 x\!-\!5)
  \nonumber \\
  &\!-\!& x^2 \Bigl\{\!\big\{\{Q [15 \!+\! (x\!-\!12) x]\} \!+\! x [x (23
 \!+\! x)\!-\!8]\big\} \!+\!
 e^2 x \{10 Q (x\!-\!1) \!+\! x [x (37\!+\! x)\!-\!16 ]\}\!\Bigr\} \nonumber \\
 &\!+\!& 2 e^2 x \{e^4 (x\!-\!1) \!-\! 2 e^2 (x\!-\!1) x \!+\!
 x^3 [4 \!+\! x (3 \!+\! x)]\} \epsilon^2\biggr\}
 +  a^4 x^3 \biggl\{10 e^8 \!+\! x^3 \{Q [27 \nonumber \\
 &\!+\!& x (5 x\!-\!28)] \!-\! 5 x [x (15 \!+\! x)\!-\!12]\} \!+\!
 e^6 x [\!-\!65 \!+\! 29 x \!+\! 10 (x\!-\!1) \epsilon^2] -
 e^4 x \bigl\{4 Q (3 \!-\! 5 x) \nonumber \\
 &\!+\!& x \{155 \!-\! 123 x \!-\! 4 x^2
 \!+\! [29 \!+\! x (5 x\!-\!14)] \epsilon^2\}\bigr\} -
 e^2 x^2 \Bigl\{4 Q [9 \!+\! (x\!-\!12) x] \!+\!
 x \bigl\{ \{x [10 \nonumber \\
 &\!+\!& x (13  \!+\! 10 x)]\!-\!21\} \epsilon^2+ 160
 \!-\! 169 x \!-\! 9 x^2\bigr\}\Bigr\}\biggr\} + a^2 x^5 \bigl\{12 e^8 \!+\!
 x^3 \{Q [9 \!+\! x (7 x\!-\!36)] \nonumber \\
 &\!+\!& x [108  \!-\! x (69 \!+\! 7 x)]\}
 \!+\! e^2 x^2 \{Q [(54  \!-\! 5 x) x\!-\!21]
 \!+\! 2 x [x (65 \!+\! 6 x)\!-\!126] \nonumber \\
 &\!+\!& x [x^2 (16 x\!-\!13)\!-\!27]
 \epsilon^2\} \!+\! e^4 x \{4 Q (4 \!-\! 5 x) \!+\! 219 x  \!+\!
 x [4 (9 \!+\! x (5 x\!-\!6)) \epsilon^2 \nonumber \\
 &\!-\!& 5 x (16 \!+\! x)]\}
 \!-\! 4 e^6 \{Q \!+\! x [3 (7 \!+\! \epsilon^2)\!-\!4 x (1 \!+\!
 \epsilon^2)]\}\bigr\}\!\Biggr\}, \\
\alpha_4&=& \!-\!16 e x^7 \Bigl\{2 x^5 [2 e^2 \!+\! (x\!-\!3) x]^2
\!+\! a^4 \bigl\{4 e^4 (x\!-\!1) \!-\! 8 e^2 (x\!-\!1) x \!+\!
 x^2 \{\!-\!3 \!+\! x [7 \!+\! x (3 \!+\! x)]\}\bigr\} \nonumber \\
 &-& a^2 x^2 \bigl\{12 e^4 (x\!-\!1) \!+\! 4 e^2 (x\!-\!3)^2 x \!+\!
 x^2 \{\!-\!27 \!+\! x [9 \!+\! x (3 x\!-\!1)]\}\bigr\}\Bigr\}
 \epsilon, \\
\alpha_5&=& 16 a x^7 [a^2 (e^2 \!-\! x) \!+\! (2 e^2 \!-\! 3 x) x^2]
 \{4 a^2 (e^2 \!-\! x) \!+\! [2 e^2 \!+\! (x\!-\!3) x]^2\}, \\
\alpha_6&=& 32 \Bigl\{a^2 x^2[a^2(x\!-\!e^2) \!+\! x^2 (3 x\!-\!2
 e^2)]^2 \{x^4 \!+\! a^2 [x (2 \!+\! x)\!-\!e^2]\} \{Q (x\!-\!1) \!+\!
 x [a^2 \!+\! e^2 \nonumber \\
 &\!+\!& x (2 x\!-\!3)]\}
 \!+\! a^2 [2 x^3 \!+\! a^2 (1 \!+\! x)] \{x^4 \!+\! a^2 [x (2 \!+\! x)\!-\!e^2]\}^2
 \{e^2 x^2\epsilon^2-[a^2 \!+\! e^2  \nonumber \\
 &\!+\!& (x\!-\!2) x](Q \!+\! x^2)\}
  [2 x^3 \!+\! a^2 (1 \!+\! x)]^2 \{x^4 \!+\! a^2 [x (2 \!+\!
 x)\!-\!e^2]\} \{2 a^2 (e^2 \!-\! 2 x) [a^2 \!+\! e^2 \nonumber \\
 &\!+\!& (x\!-\!2) x](Q \!+\! x^2)
 \!-\! e^2 x^2 [(x\!-\!1) x^4 \!+\! a^4 (1 \!+\! x) \!+\!
 2 a^2 (e^2 \!-\! 2 x \!+\! x^3)] \epsilon^2\}
 + [a^2 \!+\! e^2 \!+\! (x \nonumber \\
 &\!-\!& 2) x] [2 x^3  \!+\! a^2 (1
 \!+\! x)]^3\bigl\{e^2\epsilon^2 x^2 \{a^2 [e^2 \!-\! x (2 \!+\! x)]\!-\!x^4\}
 \!-\![a^2 (2 x\!-\!e^2)^2 (Q \!+\! x^2)]\bigr\}\Bigr\}.
\end{eqnarray}
The corresponding expressions for $L_i$ and $E_i$ for non charged
particles ($\epsilon=0$) and photons ($\mu=0$) are much simpler
and shorter \cite{dok11}.

{\sl Acknowledgements.} This research was supported in part by the
Russian Foundation for Basic Research grant No. 10-02-00635.

\end{document}